\documentclass[aps,prd,floats,preprintnumbers,showpacs,nofootinbib,superscriptaddress]{revtex4}
\usepackage{epsfig}
\usepackage{amsmath}
\usepackage{amssymb}
\usepackage{amsthm}
\usepackage{color}
\usepackage{slashed}
\usepackage{subfigure}
\usepackage{multirow}
\usepackage{longtable}
\usepackage{comment}

\begin{document}

\preprint{ NUHEP-TH/14-04}

\title{Neutrino Masses, Grand Unification, and Baryon Number Violation}

\author{Andr\'e de Gouv\^ea}
\affiliation{Northwestern University, Department of Physics \& Astronomy, 2145 Sheridan Road, Evanston, IL~60208, USA}

\author{Juan Herrero-Garc\'{\i}a}
\affiliation{Northwestern University, Department of Physics \& Astronomy, 2145 Sheridan Road, Evanston, IL~60208, USA}
\affiliation{Dept.~de F\'{\i}sica Te\'orica and IFIC, Universidad de Valencia-CSIC, Ed.~de Institutos de Paterna, Apt.~22085, 46071, Valencia, Spain}

\author{Andrew Kobach}
\affiliation{Northwestern University, Department of Physics \& Astronomy, 2145 Sheridan Road, Evanston, IL~60208, USA}


\begin{abstract}

If grand unification is real, searches for baryon-number violation should be included on the list of observables that may reveal information regarding the origin of neutrino masses. Making use of an effective-operator approach and assuming that nature is $SU(5)$ invariant at very short distances, we estimate the consequences of different scenarios that lead to light Majorana neutrinos for low-energy phenomena that violate baryon number minus lepton number ($B-L$) by two (or more) units, including neutron--antineutron oscillations and $B-L$ violating nucleon decays. We find that, among all possible effective theories of lepton-number violation that lead to nonzero neutrino masses, only a subset is, broadly speaking, consistent with grand unification. 

\end{abstract}

\maketitle

\section{Introduction}

While nonzero neutrino masses were observed over a decade ago, the mechanism that leads to them remains elusive \cite{deGouvea:2013onf}. Several qualitatively different and equally compelling theoretical options are consistent with all available neutrino data. Hope for significant progress relies heavily on data from a variety of upcoming observations and experiments.     

A popular option for rendering the neutrinos massive is to consider that lepton number is violated, spontaneously or explicitly, at some new energy scale. If the effective new-physics scale is higher than the electroweak symmetry-breaking scale $v=174$~GeV, neutrino masses are predicted to be nonzero and parametrically smaller than the known fundamental charged-fermion masses, in agreement with observations. Furthermore, regardless of how lepton number is violated in the ultraviolet, one generically expects neutrinos to be Majorana fermions, a fact that would be revealed by the observation of neutrinoless double-beta decay \cite{deGouvea:2013zba,deGouvea:2013onf}. It is unlikely, however, that the observation of neutrinoless double-beta decay, even when combined with precise information from neutrino oscillations, will reveal the details of how lepton number is broken. Such information, if at all accessible, will rely on other observations, including searches for charged-lepton flavor violation, new degrees of freedom at the TeV scale, and the violation of CP-invariance. We advocate that, if grand unification is a reality, searches for baryon-number violation should be included in the list of observables that may reveal information regarding the origin of neutrino masses. 

In the absence of new light (masses below the weak scale) degrees of freedom, nonzero neutrino masses are entirely captured by the dimension-five Weinberg operator \cite{Weinberg:1979sa}:\footnote{For future convenience, we use the operator-numbering scheme from Refs.~\cite{Babu:2001ex,de Gouvea:2007xp}.} 
\begin{equation}
{\cal O}_1 = \frac{(HL)(HL)}{\Lambda} + \text{h.c.}
\label{eq:dim5}
\end{equation}
 where $L$ is a lepton doublet, $H$ is the Higgs doublet, and flavor indices have been omitted. $\Lambda$ is the effective mass scale of the the operator. Experimental information on neutrino masses translates into $\Lambda\sim 10^{14-15}$~GeV.
 
Most other consequences of the physics behind lepton-number violation are not captured by Eq.~(\ref{eq:dim5}). Other lepton-number-violating (LNV) consequences of different would-be new physics responsible for nonzero Majorana neutrino masses can be realized if one considers different effective operators, of dimension higher than five, that mediate lepton-number violation. Relations among the different effective operators are, of course, model dependent. One can explore several different classes of models by assuming that a specific  effective operator of mass-dimension seven or higher captures the ``leading'' effects of the new physics, and then assume that the Weinberg operator is related to the leading one through quantum corrections \cite{Babu:2001ex,de Gouvea:2007xp,Angel:2012ug}. The strategy is as follows. Assume the existence of new heavy fields that couple in such a way that lepton number is broken. Integrating out the new heavy fields at the tree-level will lead to a set of higher dimensional operators (the leading ones). At the same time, when integrated out at the loop-level, the same fields will yield Eq.~(\ref{eq:dim5}). The two operators are hence related to one another (for concrete models, see, for example, Refs.~\cite{Babu:2001ex,de Gouvea:2007xp,Angel:2012ug,delAguila:2012nu,Angel:2013hla}, building on the legacy of the pioneering work of \cite{Zee:1980ai,Zee:1985rj,Babu:1988ki}, and for more on the LNV effective operator approach to neutrino masses and neutrinoless double-beta decay, see Ref.~\cite{delAguila:2012nu}). It is possible to estimate, without specifying the details of the physics that led to the leading operator, the coefficient of the Weinberg operator as a function of the effective scale of the leading operator and known Standard Model parameters (e.g., gauge or Yukawa couplings) up to order-one constants \cite{de Gouvea:2007xp,Angel:2012ug}. The effective-operator approach allows one to analyze broad classes of models, including order-of-magnitude estimates of low-energy phenomenological consequences of the new physics, without the requirement that one constructs ultraviolet-complete Lagrangians. For most scenarios, the physics scale associated to the new physics is much smaller than $10^{14}$~GeV. Indeed, in some instances, the masses of the new particles are constrained to be at most a few TeV, even if one requires all new coupling to be of order one.

The fermion content of the Standard Model, the serendipity of gauge anomaly cancellations, and the fact that the gauge couplings seem to unify at high energies can all be accounted for if one assumes that, at very high energies, nature can be described by a spontaneously broken grand-unified gauge theory (GUT). If nature is supersymmetric above the TeV scale, the GUT scale is inferred to be around $10^{16}$~GeV. As far as this work is concerned, it suffices to appreciate that the GUT scale is higher than the scale of lepton-number breaking, regardless of the details of the LNV sector, if the LNV physics is responsible for the observed nonzero neutrino masses. 

GUTs imply that lepton-number violation is intimately related to baryon-number violation. In GUTs, the different standard-model fermion multiplets are interpreted as different components of GUT multiplets in such a way that neither lepton number nor baryon number can be successfully assigned to any of the matter multiplets.\footnote{As is well known, conserved global symmetry charges related to baryon number minus lepton number can, depending on the particle content of the theory, be assigned to the different GUT matter multiplets.} If the GUT hypothesis is correct, the same physics that leads to lepton-number violation and nonzero neutrino masses also leads, necessarily, to baryon-number violation and, more precisely, to $B-L$ violation. For concrete scenarios that explore this relation see, for example, Refs.~\cite{Chang:1980ey,Chacko:1998td,Babu:2001qr,Dutta:2005af,Mohapatra:2006ei,Mohapatra:2009wp,Babu:2012vc,Babu:2013jba,Babu:2013yww}. Here we explore this connection by establishing, at the order-of-magnitude level, expected rates for different baryon-number-violating (BNV) processes as functions of what is known about neutrino masses, without making reference to specific ultraviolet-complete models. 

We will restrict our discussion to $SU(5)$ GUTs. For larger gauge groups, the matter content of the theory is larger than that of the Standard Model, and we choose not to consider the hypothesis that there are more ``light'' degrees of freedom. As a concrete example, in the case of $SO(10)$ GUTs, $U(1)_{B-L}$ is a gauged subgroup of $SO(10)$ so $B-L$ violating effects are entwined with GUT-breaking effects. Furthermore, in $SO(10)$ GUTs, one predicts the existence of standard-model gauge-singlet fermions (right-handed neutrinos). Whether these states have GUT-like masses, intermediate masses, or survive down to the weak scale and beyond depends on the details of GUT-breaking and the field-content at the GUT-scale, and we do not address this very interesting issue here. On the other hand, by choosing an $SU(5)$ route, we are implicitly addressing larger gauge groups, assuming that they are broken in such a way that an effective ``intermediate'' $SU(5)$ description is appropriate. If this is the case, all non-$SU(5)$ new fields are considered to have GUT-masses and safely decouple from the discussion.  

\section{Dimension Five}

In order to explain our strategy, we first discuss the scenarios that lead to the Weinberg operator at the tree-level. It is well known that there are three ways to reach ${\cal O}_1$ at the tree-level \cite{Ma:1998dn}: integrating out a Standard Model gauge singlet fermion (Type-I seesaw), an $SU(2)_L$ triplet scalar with zero hypercharge (Type-II seesaw), or an $SU(2)_L$ triplet fermion with zero hypercharge (Type-III seesaw) \cite{seesaw}. If the $SU(5)$-GUT hypothesis is correct, these different fields also need to be interpreted as components of $SU(5)$ multiplets, henceforth referred to, generically, as $X_5$. In the case of the $SU(2)_L$ gauge singlet fermions, these ``fit'' inside $SU(5)$ gauge singlets or be a part of $SU(5)$  $\mathbf{24}$s, while the $SU(2)_L$ triplet fermions would be part of an $SU(5)$  $\mathbf{24}$, and the triplet scalar would be part of a $\mathbf{15}$. Since the process of integrating out a subset of the fields in $X_5$ leads to $|\Delta L|=2$ effects, the process of integrating out all components of $X_5$ will lead to many different manifestations of $|\Delta(B-L)|=2$ at low energies. 

If $SU(5)$ were not broken, the act of integrating out $X_5$ would lead to the effective operator
\begin{equation}
{\cal O}^{\rm GUT}_1 = \frac{(\Phi^i\psi^{\dagger}_i)(\Phi^i\psi^{\dagger}_i)}{\Lambda} + \text{h.c.},
\label{eq:dim5_GUT}
\end{equation}
where $\psi$ is a matter field that transforms as a $\mathbf{\bar{5}}$ under $SU(5)$ and $\Phi$ is a scalar multiplet that also transforms as a $\mathbf{\bar{5}}$:
\begin{eqnarray}
\psi^i &=& \left( \begin{array}{ccccc} d_r^c & d_b^c & d_g^c & e & -\nu \end{array} \right), \\
\Phi^i &=& \left( \begin{array}{ccccc} \phi_r & \phi_b & \phi_g & \phi^- & \phi_0^{\dagger} \end{array} \right),
\end{eqnarray}
$i=1,2,3,4,5$. Here, $d^c_n~(n=r,g,b), e, \nu$ are, respectively, left-handed down-type antiquark fields of different colors, left-handed charged-lepton fields, and left-handed neutrino fields. Flavor indices have been omitted. The colored scalars $\phi_n~(n=r,g,b)$ are the components of the color-triplet Higgs field ${\mathbf H}_c$, while $\phi^+$ and $\phi^0$ are the components of the Standard Model Higgs field,
\begin{eqnarray}
H^\alpha &=& \left( \begin{array}{cc} \phi^+ & \phi_0 \end{array} \right), \hspace{0.25in} \tilde{H}^\alpha = \left( \begin{array}{cc} \phi_0^\dagger & -\phi^- \end{array} \right).
\end{eqnarray}
The remaining Standard Model matter fields are part of a multiplet $\chi$, which transforms as a $\mathbf{10}$ under $SU(5)$: 
\begin{eqnarray}
\chi_{ij} = \frac{1}{\sqrt{2}}\left( \begin{array}{ccccc} 
0 & u_g^c & -u_b^c & u_{r} & d_{r}    \\
-u_g^c & 0 & u_r^c & u_{b} & d_{b}  \\
u_b^c & - u_r^c & 0 & u_{g} & d_{g}  \\
-u_{r} & -u_{b} & -u_{g} & 0 & e^c  \\
- d_{r} & -d_{b} & -d_{g} & -e^c & 0
\end{array} \right),
\end{eqnarray}
$i,j=1,2,3,4,5$ and $\chi_{ij}=-\chi_{ji}$. Here $u^c_n,u_n,d_n~(n=r,g,b)$ are, respectively, left-handed up-type antiquark fields, left-handed up-type quark fields and left-handed down-type quark fields of different colors, and $e^c$ are left-handed charged-antilepton fields. Flavor indices have been omitted.

Eq.~(\ref{eq:dim5_GUT}) contains not only the Weinberg operator, Eq.~(\ref{eq:dim5}), but also operators that violate baryon number and lepton number. The color-triplet Higgs $\mathbf{H}_c$ cannot be assigned lepton number or baryon number, but it can be assigned $B-L$ in such a way that all Yukawa interactions preserve $B-L$.\footnote{This is well known. At the renormalizable level, minimal $SU(5)$ conserves a global $U(1)_X$ if one assigns nontrivial charges to $\chi$, $\psi$ and $\Phi$. $B-L$ charges are a combination of $U(1)_X$ and electroweak hypercharge.} It is easy to see that $\mathbf{H}_c$ has $B-L$ charge $+2/3$ so the components of Eq.~(\ref{eq:dim5}) $\propto (d^{c\dagger}\mathbf{H}_c)^2$) or $\propto (d^c\mathbf{H}_c^{\dagger})(LH)$ violate, at the tree-level and after $\mathbf{H}_c$ decay, either baryon number by plus or minus two units or lepton number by minus one unit and baryon number by one unit (or vice versa). If $SU(5)\to SU(3)_c\times SU(2)_L\times U(1)_Y$ breaking occurred after the $X_5$ fields were integrated out, the coefficients of these different $\Delta(B-L)=\pm2$ operators would be the same, modulo renormalization group running effects between the GUT-scale and the infrared.  

If grand unification is real and neutrino masses are a consequence of LNV new physics, GUT-breaking occurs at an energy scale that is higher than that of the lepton-number breaking. Hence, below the GUT-breaking scale, the masses and couplings of the various components of $X_5$ are different from one another, and the coefficients of the Standard Model effective operators in Eq.~(\ref{eq:dim5_GUT}) are no longer the same, at any energy scale.\footnote{Another important issue is that the colored Higgs triplet fields, which violate $B+L$ and mediate nucleon decay, may have masses which are larger than those of the $X_5$ fields, rendering the effective operators that contain $\mathbf{H}_c$ inconsistent from an effective field theory point of view. Experimentally, their masses are bounded to be above $10^{12}$~GeV \cite{Golowich:1981sb,Dorsner:2012nq,Dorsner:2012uz}. In order to illustrate our strategy we ignore this fact.}

Nonetheless, the BNV operators are still present, and their effective couplings are still related to those of the Weinberg operator, and the latter are still constrained by the low-energy observations related to neutrino masses. The relationship is just more model dependent. For example, the $SU(5)$ version of the Type-I, Type-II, or Type-III seesaw will lead to different relative coefficients among the components of  Eq.~(\ref{eq:dim5_GUT}), and different manifestations of GUT breaking also lead to different relations. 

If one were to assume that the GUT-breaking effects are not especially dramatic, i.e., that the several $|\Delta (B-L)|= 2$ operators have effective couplings of the same order of magnitude, one would be able to relate, at the order-of-magnitude level, LNV physics to BNV observables. This should allow one to relate, semi-quantitatively, neutrino masses to the rates for different $B-L$ processes. The comparison can be done at the effective-operator level, i.e., one can establish a relationship between possible new physics that leads to nonzero neutrino masses and the rates of several BNV observables. We first identify all ${\cal O}^{\rm GUT}_J$ ($J=1,2,3,\ldots$) that contain leading operators ${\cal O}_I$ ($I=1, 2, 3,\ldots$), following the numbering scheme introduced in Ref.~\cite{Babu:2001ex}. We then identify all the other components of ${\cal O}^{\rm GUT}_J$ which, as will be shown later, all mediate $B-L$ violation by an even number of units. Using the procedure carried out in Ref.~\cite{de Gouvea:2007xp}, we identify the value of the effective scale $\Lambda$ of the operator ${\cal O}_J^{\rm GUT}$ with the one required to ``explain'' the neutrino masses via ${\cal O}_I$. We finally use $\Lambda$ in order to estimate the rate of the different BNV processes mediated by ${\cal O}_J^{\rm GUT}$.

In the case of a dimension-five leading operator, Eq.~(\ref{eq:dim5_GUT}) contains Eq.~(\ref{eq:dim5}) and neutrino masses require $\Lambda\sim 10^{14-15}$~GeV. Further taking into account that the $\mathbf{H}_c$ must also be integrated out and that their masses are also of order the GUT scale, we conclude that all baryon-number violating phenomena mediated by this physics are very safely outside the reach of BNV probes. In the next sections, we show that this is certainly not the case for most other leading effective operators. 

\section{$\Delta B$, $\Delta L$, and Operator Dimensions}
\label{sec:def_ops}

We will only consider higher-dimensional operators constructed out of the matter fields $\psi$ and $\chi$, and the Higgs field $\Phi$, as defined in the previous section. As in Refs.~\cite{Babu:2001ex,de Gouvea:2007xp}, we do not consider operators with derivatives and gauge boson field strengths. Finally we are interested in operators that contain the $|\Delta L|=2$ operators ${\cal O}_I$ defined and described in Refs.~\cite{Babu:2001ex,de Gouvea:2007xp}.

It is interesting to explore what, if any, are the relations between an operator's mass-dimension and its total lepton and baryon-number charges. Very generically, we arrive at the following relationship between $\Delta L$, the lepton number of a given operator, $\Delta B$, the baryon number of a given operator, and $D$, the mass-dimension of the operator. Note that we do not allow for operators that contain the very heavy color-triplet Higgs fields $\mathbf{H}_c$.
\begin{equation}
\label{bldim}
\left| \frac{1}{2}   \Delta B + \frac{3}{2}   \Delta L\right| \in \mathbb{N}\left\{ \begin{array}{ll} \text{odd} &\leftrightarrow \text{$D$ is odd,} \\ \text{even} &\leftrightarrow \text{$D$ is even.} \end{array} \right.
\end{equation}
We refer readers to Appendix~\ref{operatordim} for the details of the derivation and other information. See also Ref.~\cite{Rao:1983sd}. Eq.~(\ref{bldim}) assumes only hypercharge and Lorentz invariance, and it remains unchanged if right-handed neutrinos with lepton number $+1$ are added to the Standard Model. We spell out some of its consequences below, concentrating on the ones that are most relevant to the subject at hand.  
\begin{itemize}
\item Operators with $|\Delta L|=2$, $\Delta B=0$ have odd mass dimension. The lowest such operator is dimension five (Eq.~(\ref{eq:dim5})).
\item Since $\Delta B/2 + 3\Delta L/2\equiv \Delta (B-L)/2 + 2(\Delta L)\equiv \Delta (B+L)/2 + (\Delta L)$ is an integer, $|\Delta(B-L)|$ and $|\Delta(B+L)|$ must be even numbers for any operator. Clearly, operators with odd $|\Delta B|$ also have odd $|\Delta L|$ (and vice versa). 
\item Operators with odd mass-dimension must have non-zero $\Delta B$ or $\Delta L$. In more detail, it is easy to show that, for operators with odd mass-dimension, $|\Delta(B-L)|$ is an even number not divisible by four (2, 6, 10, \ldots). All odd-dimensional operators violate $B-L$ by at least two units. For operators with even mass-dimension, $|\Delta(B-L)|$ is a multiple of four, including zero (0, 4, 8, 12, \ldots).
\end{itemize} 
As far as our discussion is concerned, we will be interested in $\mathcal{O}^{\text{GUT}}_J$ with odd mass-dimension, since all  $\mathcal{O}_I$ are odd-dimensional. Furthermore, all other components of  $\mathcal{O}^{\rm GUT}_J$ will, necessarily, violate baryon number or lepton number. It is important to emphasize that the effects discussed here are qualitatively different from many other more widely known BNV effects from GUTs. As is well known, $SU(5)$ GUTs have a global $U(1)_X$ symmetry which is proportional to $B-L$.  If $U(1)_X$ is conserved, this implies that $\Delta(B-L)=0$. Our result implies that all effects that conserve $U(1)_X$, including, for example, the exchange of heavy GUT gauge bosons or Higgs fields, lead to effective operators that have even mass-dimension. The lowest-energy BNV effective operators, for example, are dimension six and conserve $B-L$ \cite{Weinberg:1979sa,Wilczek:1979hc} (including $QQQL$ and $u^cu^cd^ce^c$) and, naively, are not related to the physics responsible for nonzero Majorana neutrino masses.

We list and number all odd-dimensional $\mathcal{O}^{\rm GUT}_J$, $J=1,2,3,\ldots$, with mass-dimension $D\le 9$ in Table~\ref{app_table}.
%
%\section{Odd-Dimensional GUT Effective Operators ($D\le 9$)}
%
\begin{table}[t]
\caption{$SU(5)$ GUT-invariant effective operators that consist of the matter fields $\psi,\chi$ or Higgs boson field $\Phi$ and have odd mass-dimension $D=5,7,9$. We exclude operators with derivatives and gauge boson field strengths, and ignore operators that consist of the simple `composition' of two lower-dimensional gauge-invariant operators. 
%(say, a dimension-four `Yukawa coupling' and ${\cal O}^{\rm GUT}_1$, or the dimension-two $\Phi\Phi^{\dagger}$ ``Higgs mass'' and some other ${\cal O}^{\rm GUT}_J$).  
The operators $\mathcal{O}_I$ contained within $\mathcal{O}_J^\text{GUT}$ are listed according to the numbering scheme from Refs.~\cite{Babu:2001ex,de Gouvea:2007xp}, using bold-face font for the $\mathcal{O}_I$ operator that gives the dominant contribution to neutrino masses. The `singlets' operator refers to $e^c e^c u^c u^c d^{c\dagger}d^{c\dagger}$, not present in Refs.~\cite{Babu:2001ex,de Gouvea:2007xp}. The star symbol indicates operators that would vanish if one did not take into account the existence of more than one generation of fermions. See text for details. 
} \label{app_table}
\begin{tabular}{ | c | c | c | c |}
\hline
Dimension &$J$, for $\mathcal{O}_J^\text{GUT}$ & Operator & $I$, for $\mathcal{O}_I$   \\ \hline \hline 
5&$1$ & $\psi^i \Phi^\dagger_i   \psi^j \Phi^\dagger_j $ &  {\bf1} \\ \hline
7&$2_a$ &  $\epsilon_{ijklm}  \chi^{\dagger ij} \chi^{\dagger kl} \psi^m \psi^n \Phi^\dagger_n$  & ${\bf4_a}$, 8    \\
7&$2_b^\star$ &  $\epsilon_{ijklm}  \chi^{\dagger ij} \psi^k \psi^l \chi^{\dagger mn}\Phi_n^\dagger$ & ${\bf 4_b}$, $8$    \\
7&3 &  $\chi_{ij} \psi^i \psi^j \psi^k  \Phi_k^\dagger$  &  2, ${\bf 3_b}$  \\ 
7&$4^\star$ &  $\epsilon_{ijklm} \psi^i \psi^j \psi^k \psi^l \Phi^m$  & \\ \hline
9&5 & $\chi_{ij} \chi_{kl} \psi^i \psi^j \psi^k \psi^l$ &  9, 10, ${\bf11_b}$ \\
9&$6_a$ & $\epsilon_{ijklm} \psi^i \chi^{\dagger jk} \chi^{\dagger lm} \chi_{no} \psi^n \psi^o$  & $13$, ${\bf 14_b}$, $16$, $19$ \\ 
9&$6_b$ & $\epsilon_{ijklm} \psi^i \psi^j \chi^{\dagger kl} \chi^{\dagger mn} \chi_{no} \psi^o$  & $13$, ${\bf 14_b}$, $16$, $18$, $19$ \\
9&$6_c^\star$ &  $\epsilon_{ijklm}  \psi^i \psi^j \psi^k \chi^{\dagger lm} \chi^{\dagger no} \chi_{no} $ & ${\bf 14_a}$, $16$, $18$ \\
9&$7_a$ & $\epsilon_{ijklm} \epsilon_{nopqr} \psi^i \chi^{\dagger jk} \chi^{\dagger lm} \psi^n \chi^{\dagger op} \chi^{\dagger qr}$  & ${\bf 12_a}$, $20$, singlets \\
9&$7_b^\star$ & $\epsilon_{ijklm} \epsilon_{nopqr} \psi^i \psi^j \chi^{\dagger kl} \chi^{\dagger mn} \chi^{\dagger op} \chi^{\dagger qr}$  & ${\bf 12_b}$, $20$, singlets \\
9&$8_a^\star$ &  $\epsilon_{ijklm}  \psi^i  \psi^j \psi^k \psi^l \chi^{\dagger mn} \psi^\dagger_n $ & ${\bf 17}$  \\
9&$8_b^\star$ &  $\epsilon_{ijklm}  \psi^i  \psi^j \psi^k \chi^{\dagger lm} \psi^n \psi^\dagger_n  $ &  ${\bf 15}$, $17$  \\
9&9 &  $\epsilon_{ijklm}  \psi^i  \Phi^j \chi^{\dagger kl} \chi^{\dagger mn} \Phi^\dagger_n \psi^o \Phi^\dagger_o $&  {\bf 6}   \\
\hline
\end{tabular}
\end{table}

\section{Dimension Seven}
\label{sec:seven}

The dimension-seven operators one can construct out of $\psi$, $\chi$ and $\Phi$, listed in Table~\ref{app_table}, are $\mathcal{O}^{\rm GUT}_{2_a,2_b,3,4}$. When expressed in terms of their Standard Model ``components,'' all related operators violate $B-L$ by two units (violating $B-L$ by six units requires operators of much higher mass-dimension). Since their mass-dimension is too small ($|\Delta B|=2$ operators are at least dimension-nine), all operators either violate lepton number by two units, or baryon number by one unit and lepton number by minus one unit (or vice versa).   

Operator $\mathcal{O}^{\rm GUT}_{2_a}$ contains the following ``components'':
\begin{equation}
\frac{1}{\Lambda^3}\epsilon^{ijklm}  (\chi_{ij} \chi_{kl})( \psi_m^\dagger \psi_n^\dagger) \Phi^n \supset \left\{\frac{\epsilon^{\alpha\beta}}{\Lambda^3}\epsilon_{\delta\gamma}H^*_\alpha (L^\dagger_\beta d^{c\dagger}) (Q^\gamma Q^\delta),~ \frac{\epsilon^{\alpha\beta}}{\Lambda^3}H^*_\alpha (L^\dagger_\beta d^{c\dagger})(e^c u^c),~ \frac{\epsilon^{\alpha\beta}}{\Lambda^3}\delta_{\delta\gamma} H^*_\alpha  (L^\dagger_\beta L^\dagger_\delta) (Q^\gamma u^c) \right\},
\label{eq:O2}
\end{equation} 
where we have omitted terms that contain the color-triplet Higgs fields, which we henceforth assume are GUT-scale heavy, and omitted flavor indices. $SU(3)$ contractions are implicit, but we have made the $SU(2)$ contractions explicit. Our notation does not address the possibility of forming operators with a different Lorentz structure using $\sigma^\mu$, $\overline{\sigma}^\mu$, $\sigma^{\mu\nu}$ or $\overline{\sigma}^{\mu\nu}$. As will become clear later, these (sometimes distinct) operators are expected to lead to the same results as the operators above up to, at most, order-one corrections. The act of integrating out the color-triplet Higgs fields will lead to higher-dimensional BNV and LNV operators, which we expect are subleading when compared to the effects we are discussing here.

The first term on the right-hand side of Eq.~(\ref{eq:O2}) has $\Delta B = -\Delta L = 1$, while the other two terms violate lepton number by two units. In more detail, the second term is operator ${\cal O}_8$ in Refs.~\cite{Babu:2001ex,de Gouvea:2007xp}, while the third term is ${\cal O}_{4a}$ in Refs.~\cite{Babu:2001ex,de Gouvea:2007xp}.\footnote{Henceforth, ${\cal O}_{I}$ for $I=1,2,3,\ldots$ will refer to $|\Delta L|=2$ operators tabulated in Refs.~\cite{Babu:2001ex,de Gouvea:2007xp}, while ${\cal O}^{\rm GUT}_{J}$ for $J=1,2,3,\ldots$ will refer to the operators tabulated in Table~\ref{app_table}.} This information is included in the third column of Table~\ref{app_table}.

Using the results of Refs.~\cite{de Gouvea:2007xp,Angel:2012ug}, we estimate that, for ${\cal O}_{4a}$, the neutrino data require $\Lambda \sim 4\times 10^{12}$~GeV, while for ${\cal O}_{8}$, neutrino data call for $\Lambda \sim 6\times 10^6$~GeV. Since both are part of the same $\mathcal{O}^{\rm GUT}_{2_a}$, we will choose $\Lambda$ in Eq.~(\ref{eq:O2}) such that it agrees with the {\sl largest} of the two: $\Lambda \sim 4\times 10^{12}$~GeV. In this way, the ${\cal O}_{4a}$ component of $\mathcal{O}^{\rm GUT}_{2_a}$ ``fits'' the neutrino data, while the ${\cal O}_{8}$ provides a subdominant contribution (around six orders of magnitude smaller). This analysis indicates that, unless GUT-breaking effects are very large, grand unification implies that there are no scenarios where the neutrino masses are a consequence of physics that yield, at the tree-level, ${\cal O}_8$ -- other related GUT degrees of freedom, which `manifest themselves' as ${\cal O}_{4a}$, will contribute at a more substantive level. The fact that  ${\cal O}_{4a}$ is the dominant LNV operator in $\mathcal{O}^{\rm GUT}_{2_a}$ is indicated (using bold-face font) in Table~\ref{app_table}.

After electroweak symmetry breaking, $\epsilon_{\alpha\beta} \Lambda^{-3} H^{*\alpha} \left( L^{\dagger\beta} d^{c\dagger} \right) \left( Q^\alpha Q^\beta \right)$ mediates different nucleon-decay processes, including $p \rightarrow \pi^+ \nu$, $p \rightarrow \rho^+ \nu$, $n\to\pi^0\nu$, etc. We estimate 
\begin{equation}
\Gamma(N \rightarrow M \nu)\sim \frac{1}{8\pi} \left(\frac{v}{\Lambda^3}\right)^2 \Lambda_\text{QCD}^5,
\label{eq:N_width}
\end{equation}
where $N$ is a nucleon (proton or neutron), $M$ is a meson ($\pi, \rho, K$, etc.), and $\Lambda_{\rm QCD}\equiv 0.25$~GeV. We applied our estimation procedure to more detailed computations of the nucleon lifetime \cite{Aoki:2013yxa} and agree with them up to order-one factors. At the same order in perturbation theory, $\mathcal{O}^{\rm GUT}_{2_a}$ mediates three-body nucleon decays $N\to MM\nu$, whose contribution to the the nucleon lifetime is phase-space suppressed.

\setcounter{footnote}{0}

Given the ``neutrino mass'' value for $\Lambda$, we can estimate the nucleon lifetime. For $\Lambda \sim 4\times 10^{12}$ GeV, the lifetime of the nucleon turns out to be $\tau_{N} \sim 10^{44}$ years. To be concrete, and in order to take the many different uncertainties into account, we will henceforth present and discuss numerical estimates as follows. While relating $\Lambda$ to the neutrino masses, we impose that the neutrino masses lie between 0.05~eV and 0.5~eV, the square root of the largest confirmed neutrino mass-squared difference and the (approximate) upper bound on the sum of the neutrino masses from measurements of the large-scale structure of the universe \cite{Ade:2013zuv}, respectively.\footnote{Strictly speaking, the objects we are computing are the elements of the neutrino Majorana mass matrix. In models where the neutrino mass matrix is predicted to be hierarchical (see Ref.~\cite{de Gouvea:2007xp}), the neutrino-mass range in the text is imposed on the largest matrix elements.} The extracted $\Lambda$ values, tabulated in Table~\ref{table_bounds}, are hence uncertain by one order of magnitude, as explicitly indicated. This one order of magnitude uncertainty in $\Lambda$ translates, according to Eq.~(\ref{eq:N_width}), into a six orders of magnitude uncertainty on the nucleon lifetime. We feel that this is an appropriate way of accommodating all the approximations made here in order to estimate the nucleon lifetime given neutrino mass constraints.

\begin{table}[t]
\caption{{List of $SU(5)$ operators of mass-dimension seven or nine that violate $B-L$ by two units. We show in the third column the $\Delta L=2$ operator within the $SU(5)$ one that gives the dominant contribution to neutrino masses, together with the scale necessary for it to fit neutrino masses, in the fourth column. The predictions for the $\Delta B\neq0$ observables are listed in the last column. The notation is as follows: $N$ is a nucleon (proton $p$ or neutron $n$), $M$ is a meson ($\pi,\rho,K,\ldots$) and $\ell$ is a neutral ($\nu$) or charged ($e$) lepton. Generation indices have been omitted. See text for more details.}} 
\label{table_bounds}
{
%\small
\begin{tabular}{| c | c | c | c | c | } 
%\begin{longtable}{| c | c | c | c | c | } 
\hline
$J$ for $\mathcal{O}_J^\text{GUT}$ & Operator  & $\mathcal{O}_I$ & $\Lambda$ in GeV & Select $\Delta B \neq 0$ Observables  \\ \hline \hline
%1
%1 & $\psi^i \Phi_i^\dagger \psi^j \Phi_j^\dagger$ & ${\cal O}_1$ & $ 6\times 10^{10-11}$ & ``none'' \\ \hline
 % 2a
$2_a$ & $\epsilon_{ijklm}  \chi^{\dagger ij} \chi^{\dagger kl} \psi^m \psi^n \Phi^\dagger_n$ &  ${\cal O}_{4_a}$ & $4\times 10^{11-12}$ & $\tau(N\rightarrow M\nu ) \sim 8\pi \left( \frac{\Lambda^3}{v} \right)^2 \frac{1}{\Lambda_\text{QCD}^5} \sim 10^{37-44} $ years \\ \hline
 % 2b
{$2_b^\star$} & {$\epsilon_{ijklm}  \chi^{\dagger ij} \psi^k \psi^l \chi^{\dagger mn}\Phi_n^\dagger$} & {${\cal O}_{4_b}$} & {$6\times 10^{8-9}$} & $ \tau(N\rightarrow M\ell ) \sim 8\pi \left( \frac{\Lambda^3}{v} \right)^2 \frac{1}{\Lambda_\text{QCD}^5} \sim 10^{20-27} $ years \\  \hline
% 3
{$3$} & {$\chi_{ij} \psi^i \psi^j \psi^k  \Phi_k^\dagger$} & {${\cal O}_{3_b}$} & {$1\times 10^{10-11}$} & $\tau(N\rightarrow M\nu ) \sim 8\pi \left( \frac{\Lambda^3}{v} \right)^2 \frac{1}{\Lambda_\text{QCD}^5} \sim 10^{28-34} $ years \\ \hline
 %4
4 & $\epsilon_{ijklm} \psi^i \psi^j \psi^k \psi^l \Phi^m$ & none &  no estimate & no estimate \\ \hline
% 5
\multirow{5}{*}{5} & \multirow{5}{*}{$\chi_{ij}\chi_{kl} \psi^i \psi^j \psi^k \psi^l$} & \multirow{5}{*}{${\cal O}_{11_b}$} & \multirow{5}{*}{$2\times 10^{6-7}$} & 
$ \tau(N \rightarrow M \nu)  \sim 8\pi \left(\frac{16\pi^2}{y_d} \frac{\Lambda^3}{v} \right)^2 \frac{1}{\Lambda_\text{QCD}^5} \sim 10^{14-20}$ years \\ 
& & & & $\tau(n \rightarrow Me ) \sim 8\pi \left(\frac{16\pi^2}{y_u} \frac{\Lambda^3}{v} \right)^2 \frac{1}{\Lambda_\text{QCD}^5}  \sim 10^{10-17} $ years \\
& & & & $\tau(n-\overline{n}) \sim  \frac{\Lambda^5}{\Lambda_\text{QCD}^6}  \sim 10^{3-8}$ years \\ \hline
% 6a
\multirow{3}{*}{$6_a$} & \multirow{3}{*}{$\epsilon_{ijklm} \psi^i \chi^{\dagger jk} \chi^{\dagger lm} \chi_{no} \psi^n\psi^o$} & \multirow{3}{*}{${\cal O}_{14_b}$} & \multirow{3}{*}{$6\times 10^{7-8}$} & $ \tau(N \rightarrow M \nu) \sim 8\pi \left(\frac{16\pi^2}{y_u} \frac{\Lambda^3}{v} \right)^2 \frac{1}{\Lambda_\text{QCD}^5} \sim 10^{19-26}$ years \\ 
& & & & $\tau(n-\overline{n}) \sim  \frac{\Lambda^5}{\Lambda_\text{QCD}^6} \sim 10^{10-16}$ years \\ \hline
% 6b
\multirow{5}{*}{$6_b$} & \multirow{5}{*}{$\epsilon_{ijklm} \psi^i \psi^j \chi^{\dagger kl} \chi^{\dagger mn} \chi_{no} \psi^o$} & \multirow{5}{*}{${\cal O}_{14_b}$} & \multirow{5}{*}{$6\times 10^{7-8}$} & $ \tau(N \rightarrow M \nu)  \sim 8\pi \left(\frac{16\pi^2}{y_u} \frac{\Lambda^3}{v} \right)^2 \frac{1}{\Lambda_\text{QCD}^5} \sim 10^{19-26}$ years \\ 
& & & & $\tau(n \rightarrow Me ) \sim 8\pi \left(\frac{16\pi^2}{y_d} \frac{\Lambda^3}{v} \right)^2 \frac{1}{\Lambda_\text{QCD}^5}  \sim 10^{23-29} $ years \\
& & & & $\tau(n-\overline{n}) \sim  \frac{\Lambda^5}{\Lambda_\text{QCD}^6}  \sim 10^{10-16}$ years \\ \hline
% 6c
\multirow{3}{*}{$6_c^\star$} & \multirow{3}{*}{$\epsilon_{ijklm}  \psi^i \psi^j \psi^k \chi^{\dagger lm} \chi^{\dagger no} \chi_{no} $} & \multirow{3}{*}{${\cal O}_{14_a}$} & \multirow{3}{*}{$1\times 10^{5-6}$} & $ \tau(N \rightarrow M \nu)  \sim 8\pi \left(\frac{16\pi^2}{y_u} \frac{\Lambda^3}{v} \right)^2 \frac{1}{\Lambda_\text{QCD}^5} \sim 10^{2-9}$ years \\ 
& & & & $\tau(n \rightarrow Me ) \sim 8\pi \left(\frac{16\pi^2}{y_d} \frac{\Lambda^3}{v} \right)^2 \frac{1}{\Lambda_\text{QCD}^5}  \sim 10^{6-12} $ years \\ \hline
% 7a
\multirow{5}{*}{$7_a$} & \multirow{5}{*}{$\epsilon_{ijklm} \epsilon_{nopqr} \psi^i \chi^{\dagger jk} \chi^{\dagger lm} \psi^n \chi^{\dagger op} \chi^{\dagger qr}$} & \multirow{5}{*}{${\cal O}_{12_a}$} & \multirow{5}{*}{$2\times 10^{9-10}$} & $ \tau(p \rightarrow M \nu)  \sim 8\pi \left(\frac{16\pi^2}{y_u} \frac{\Lambda^3}{v} \right)^2 \frac{1}{\Lambda_\text{QCD}^5} \sim 10^{28-35}$ years \\ 
& & & & $\tau(n \rightarrow M \ell ) \sim 8\pi \left(\frac{16\pi^2}{y_u} \frac{\Lambda^3}{v} \right)^2 \frac{1}{\Lambda_\text{QCD}^5}  \sim 10^{28-35} $ years \\
& & & & $\tau(n-\overline{n}) \sim  \frac{\Lambda^5}{\Lambda_\text{QCD}^6} \sim 10^{18-23}$ years \\ \hline
% 7b
\multirow{3}{*}{$7_b^\star$} & \multirow{3}{*}{$\epsilon_{ijklm} \epsilon_{nopqr} \psi^i \psi^j \chi^{\dagger kl} \chi^{\dagger mn} \chi^{\dagger op} \chi^{\dagger qr}$} & \multirow{3}{*}{${\cal O}_{12_b}$} & \multirow{3}{*}{$4\times 10^{6-7}$} & $ \tau(p \rightarrow M \nu)  \sim 8\pi \left(\frac{16\pi^2}{y_u} \frac{\Lambda^3}{v} \right)^2 \frac{1}{\Lambda_\text{QCD}^5} \sim 10^{12-18}$ years \\ 
& & & & $\tau(n \rightarrow M \ell ) \sim 8\pi \left(\frac{16\pi^2}{y_u} \frac{\Lambda^3}{v} \right)^2 \frac{1}{\Lambda_\text{QCD}^5}  \sim 10^{12-18} $ years \\
\hline
% 8a
\multirow{3}{*}{$8_a^\star$} & \multirow{3}{*}{$\epsilon_{ijklm}  \psi^i  \psi^j \psi^k \psi^l \chi^{\dagger mn} \psi^\dagger_n $} & \multirow{3}{*}{${\cal O}_{17}$} & \multirow{3}{*}{$2\times 10^{2-3}$} & $ \tau(p \rightarrow M \nu)  \sim 8\pi \left(\frac{\left(16\pi^2\right)^2}{ g^2 y_d} \frac{\Lambda^3}{v} \right)^2 \frac{1}{\Lambda_\text{QCD}^5} \sim 10^{2-9}$ seconds \\ 
& & & & $\tau(n \rightarrow M e ) \sim 8\pi \left(\frac{16\pi^2}{y_d} \frac{\Lambda^3}{v} \right)^2 \frac{1}{\Lambda_\text{QCD}^5}  \sim 10^{(-3)-(+3)} $ seconds \\
\hline
% 8b
\multirow{3}{*}{$8_b^\star$} & \multirow{3}{*}{$\epsilon_{ijklm}  \psi^i  \psi^j \psi^k \chi^{\dagger lm} \psi^n \psi^\dagger_n $} & \multirow{3}{*}{${\cal O}_{15}$} & \multirow{3}{*}{$1\times 10^{5-6}$} & $ \tau(p \rightarrow M \nu)  \sim 8\pi \left(\frac{\left(16\pi^2\right)^2}{ g^2 y_d} \frac{\Lambda^3}{v} \right)^2 \frac{1}{\Lambda_\text{QCD}^5} \sim 10^{11-17}$ years \\ 
& & & & $\tau(n \rightarrow M e ) \sim 8\pi \left(\frac{16\pi^2}{y_d} \frac{\Lambda^3}{v} \right)^2 \frac{1}{\Lambda_\text{QCD}^5}  \sim 10^{6-12} $ years \\
\hline
% 9 
9 & $\epsilon_{ijklm}  \psi^i  \Phi^j \chi^{\dagger kl} \chi^{\dagger mn} \Phi^\dagger_n \psi^o \Phi^\dagger_o $ & ${\cal O}_6$ & $2\times 10^{9-10}$ & $\tau(N\rightarrow M\nu) \sim 8\pi \left(16\pi^2 \frac{\Lambda^3}{v}\right)^2 \frac{1}{\Lambda_\text{QCD}^5} \sim 10^{28-34} $ years \\ \hline
%\end{longtable}
\end{tabular}
}
\end{table}

The current most stringent experimental bound on the nucleon lifetime is $\tau(p^+ \rightarrow \rho^+ \nu) > 1.6\times10^{32}$ years, at the $90\%$ confidence level~\cite{Beringer:1900zz}, much shorter than the range indicated in Table~\ref{table_bounds} for ${\cal O}^{\rm GUT}_{2_a}$. In fact, we anticipate that, if the LNV physics information is properly ``contained'' in $\mathcal{O}^{\rm GUT}_{2_a}$, $|\Delta(B-L)|=2$ nucleon decay processes are expected to remain unobservable for the foreseeable future. We re-emphasize that we have nothing to say about  $|\Delta(B-L)|=0$ nucleon decay modes. These are mediated by physics that cannot be related in a model-independent way to the physics responsible for nonzero neutrino masses. 

Of the remaining three dimension-seven ${\cal O}^{\rm GUT}$ operators, ${\cal O}^{\rm GUT}_{2_b,3}$ contain $|\Delta L|=2$ components after GUT-breaking, as listed in Table~\ref{app_table}. ${\cal O}^{\rm GUT}_{4}$, on the other hand, only contain terms that violate both baryon-number and lepton number by one unit (in absolute value). Models that manifest themselves via  ${\cal O}^{\rm GUT}_{4}$ are not viable candidates for the physics responsible for neutrino masses.

The results in Table~\ref{table_bounds} reveal that the possibility that the physics responsible for neutrino masses is captured by ${\cal O}^{\rm GUT}_{3}$ is already constrained, but not ``ruled out'' by searches for  $p^+ \rightarrow \rho^+ \nu$. This is due to the fact that the effective scale of the operator is of order $10^{11}$~GeV, an order of magnitude lower than that of  ${\cal O}^{\rm GUT}_{2_a}$ discussed above. 

 ${\cal O}^{\rm GUT}_{2_b}$, on the other hand, requires an even smaller value of $\Lambda$ in order to ``fit'' the neutrino mass data and, in the absence of large GUT-breaking effects, is disfavored by nucleon decay searches.  ${\cal O}^{\rm GUT}_{2_b}$ has the interesting feature that it would vanish in the absence of more than one generation, hinting that, modulo a proper discussion of Yukawa couplings, some of the final-state mesons or leptons from $N\to M\ell$ are not in the first generation. As summarized in Ref.~\cite{Beringer:1900zz}, however, many nucleon decay modes of the type $N\to M\ell$ are constrained to outlive $10^{27}$~years by many orders of magnitude, including some that have second generation leptons or quarks: $\tau(n \rightarrow \rho^+ \mu^-) > 7\times10^{30}$ years, $\tau(p \rightarrow K^+ \nu) > 6.7\times10^{32}$ years, $\tau(n \rightarrow K^+ \mu^-) > 5.7\times10^{31}$, etc., all at the $90\%$ confidence level. Many three-body decay modes, like $p\to e^-\pi^+ K^+$, are are strongly constrained ($\tau(p \rightarrow e^-\pi^+K^+) > 2.5\times10^{32}$ years at the 90\% confidence level). 

In summary, if GUTs are indeed real and the physics responsible for neutrino masses is captured by an effective $|\Delta L|=2$ operator of mass-dimension seven, we conclude that 
the new physics must manifest itself via one of ${\cal O}_{3_b,4_a,4_b}$. It cannot manifest itself in the form of operators ${\cal O}_{3_a, 5, 7, 11_a}$ from Ref.~\cite{Babu:2001ex,de Gouvea:2007xp} as these are not ``present'' in ${\cal O}^{\rm GUT}_{2_a,2_b,3,4}$, and it cannot manifest itself via ${\cal O}_{8}$ since it is entwined with other $|\Delta L|=2$ operators via GUT relations. Furthermore, nucleon decay bounds disfavor  ${\cal O}^{\rm GUT}_{2_b}$ (and hence ${\cal O}_{4_b}$), while ${\cal O}^{\rm GUT}_{3}$ (and hence ${\cal O}_{2,3_b}$\footnote{The value of $\Lambda$ that allows ${\cal O}_{2,3_b}$ to accommodate the observed neutrino masses are very similar, see Ref.~\cite{de Gouvea:2007xp}.}) is already somewhat constrained by failed searches for nucleon decay, indicating that, if nature chooses to generate neutrino masses in this way, future nucleon decay searches are expected to observe baryon-number violation. ${\cal O}^{\rm GUT}_{2_a}$, which yields nonzero neutrino masses predominantly via ${\cal O}_{4_a}$, is the only scenario that is unconstrained by searches for baryon-number violation. We emphasize again that large GUT-breaking effects will allow one to evade many of the constraints. 

Before proceeding, it is instructive to quickly discuss an ultraviolet complete toy example that realizes one of the operators discussed here. At the GUT-level, we add to the Lagrangian two new scalar fields, $Z_m$ (a $\mathbf{5}$ of $SU(5)$), and $Y_{ij}$ (a $\mathbf{10}$ of $SU(5)$), along with a potential that contain the following terms
\begin{equation}
\left(h_Y\psi^i\psi^jY_{ij}+\kappa_{YZ}\Phi_i^{\dagger}Z_jY^{\dagger ij}+h_Z\epsilon^{ijklm}\chi_{ij}\chi_{kl}Z_m+\text{h.c.}\right) + M^2_Y Y_{ij}Y^{\dagger ij} + M_Z^2 Z_i Z^{\dagger i}.
\end{equation}
Integrating out the heavy fields $Y$ and $Z$ at the tree-level, ${\cal O}^{\rm GUT}_{2a}$ is generated, with $1/\Lambda^3=\kappa_{YZ}h_Y h_Z/M^2_YM^2_Z$. Integrating $Y$ and $Z$ at the one-loop level will lead to ${\cal O}^{\rm GUT}_1$, as expected. Note that the model would be invariant under $U(1)_X$ if any of the couplings $h_Y,h_Z,\kappa_{YZ}$ were to vanish. A slight variation of the model, where the $h_Z$ term is replaced by $\psi\chi Z^{\dagger}$, would lead to ${\cal O}^{\rm GUT}_{3}$.

\section{Dimension Nine}
\label{sec:nine}

The dimension-nine operators one can construct out of $\psi$, $\chi$ and $\Phi$, listed in Table~\ref{app_table}, are $\mathcal{O}^{\rm GUT}_{5,6_a,6_b,6_c,7_a,7_b,8_a,8_b}$. When expressed in terms of their Standard Model ``components,'' all related operators violate $B-L$ by two units (violating $B-L$ by six units requires operators of yet higher mass-dimension). Unlike the dimension-seven case, the dimension-nine operators also accommodate $|\Delta B|=2$ terms. 

Operator $\mathcal{O}^{\rm GUT}_{5}$ contains the following ``components:''
\begin{eqnarray} 
\frac{1}{\Lambda^5} \chi_{ij} \chi_{kl} \psi^i \psi^j \psi^k \psi^l  & \supset & \left\{  \frac{1}{\Lambda^5}e^c e^c L^i L^j L^k L^l \epsilon_{ij} \epsilon_{kl},~ \frac{1}{\Lambda^5}
d^c d^c L^i L^k Q^j Q^l \epsilon_{ij} \epsilon_{kl},~
 \frac{1}{\Lambda^5} d^c e^c L^i L^j L^k Q^l \epsilon_{ij} \epsilon_{kl}, \right. \nonumber \\
& & \left.
 \frac{1}{\Lambda^5} d^cd^cd^cd^cu^cu^c, ~
 \frac{1}{\Lambda^5} d^c d^c d^c u^c L^i Q^j \epsilon_{ij},~
 \frac{1}{\Lambda^5}d^c d^c u^c e^c L^i L^j \epsilon_{ij}
\right\}.
\label{eq:O5}
\end{eqnarray} 
As in Eq.~(\ref{eq:O2}), we omitted terms that contain the color-triplet Higgs fields and flavor indices. $SU(3)$ contractions are implicit, but we have made the $SU(2)$ contractions explicit. $\mathcal{O}^{\rm GUT}_{5}$ contains (first line of Eq.~(\ref{eq:O5})) the LNV operators ${\cal O}_9$, ${\cal O}_{10}$, and ${\cal O}_{11_b}$, which ``explain'' neutrino masses if $\Lambda\sim3\times 10^6$~GeV, $\Lambda\sim6\times 10^6$~GeV, $\Lambda\sim2\times 10^7$~GeV, respectively. As discussed in Sec.~\ref{sec:seven}, we choose the largest of these ($\Lambda\sim2\times 10^7$~GeV) for the purpose of estimating BNV observables, and conclude that grand unification disfavors LNV physics that manifests itself via ${\cal O}_9$ or ${\cal O}_{10}$. ${\cal O}_{11_b}$ is expected to dominate low-energy LNV phenomena. 

The ``operator-components'' $\Lambda^{-5}d^c d^c d^c u^c L^i Q^j \epsilon_{ij}$ and $\Lambda^{-5}d^c d^c u^c e^c L^i L^j \epsilon_{ij}$ (in the second line of  Eq.~(\ref{eq:O5})) violate baryon number by minus one unit and lepton number by one unit (and vice versa),  and mediate at low-energies nucleon decay, similar to the dimension-seven operators discussed in the previous section. The component $\Lambda^{-5}d^cd^cd^cd^cu^cu^c$ (in the second line of  Eq.~(\ref{eq:O5})), on the other hand, only violates baryon number, by two units. It gives rise to, e.g., $n-\overline{n}$ oscillations and $N+N\to{\cal M}$ inside of nuclei (where ${\cal M}$ is a state with zero baryon number), the latter proceeding via, e.g., gluon emission from one of the quark lines. 

\begin{figure}[ht]
\includegraphics[width=0.45\textwidth]{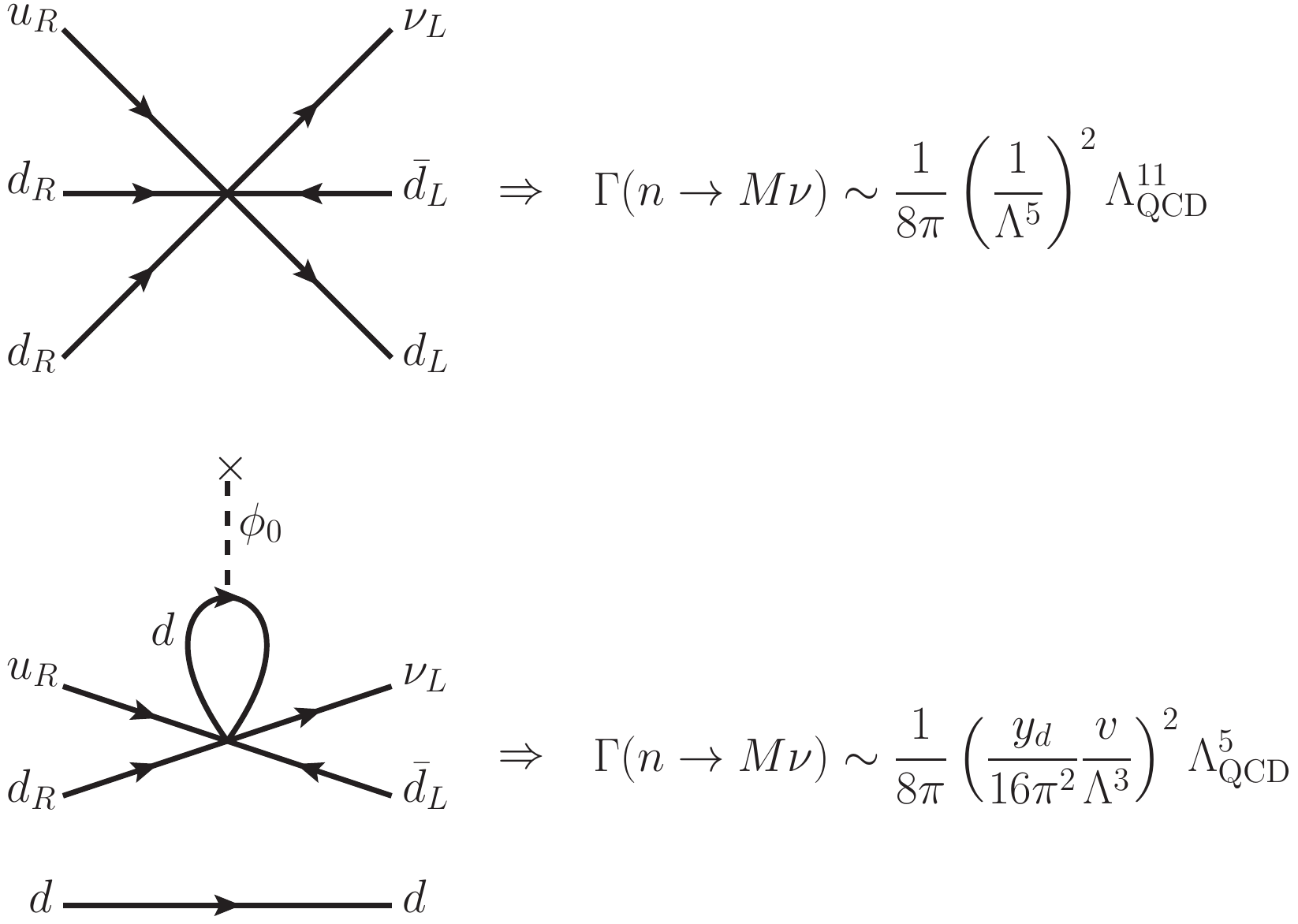}
\caption{Contributions of $\Lambda^{-5}d^c d^c d^c u^c L^i Q^j \epsilon_{ij}$ to $n\to M\nu$ at the tree-level (top), and the one-loop level (bottom), along with respective estimates for the decay width. The loop-diagram outweighs the tree-level contribution as it is related to a mass-dimension seven effective operator, as opposed to mass-dimension nine.}
\label{9to7}
\end{figure}

At the tree-level, $\Lambda^{-5}d^c d^c d^c u^c L^i Q^j \epsilon_{ij}$ and $\Lambda^{-5}d^c d^c u^c e^c L^i L^j \epsilon_{ij}$ will mediate $|\Delta(B-L)|=2$ nucleon decay processes with two and three particles in the final state, including a purely leptonic decay of the neutron: $n\to e^+e^-\nu$. At the one-loop level, by combining the known quark and charged-lepton Yukawa interactions with the physics that yields the dimension-nine operators above at the tree level, one obtains dimension-seven $|\Delta(B-L)|=2$ effective operators that mediate much faster nucleon decay processes. We estimate these effects following the procedure described in detail in Ref.~\cite{de Gouvea:2007xp}, which allowed one to relate different higher-dimensional ($D\ge 7$) $|\Delta L|=2$ leading operators to the $D=5$ Weinberg operator. For example, the tree-level operator $\Lambda^{-5}d^c d^c d^c u^c L^i Q^j \epsilon_{ij}$ can be related to one-loop dimension-seven operators that mediates $N\to M\nu$, as illustrated in Figure~\ref{9to7} in the case of $n\to M\nu$. We estimate the associated decay width as follows:
\begin{equation}
\Gamma(N \rightarrow M \nu)\sim \frac{1}{8\pi} \left(\frac{y_{d}}{16\pi^2}\right)^2 \left(\frac{v}{\Lambda^3}\right)^2 \Lambda_\text{QCD}^5,
\label{eq:N_width_9}
\end{equation}
where $y_d$ is a down-type-quark Yukawa coupling. We also illustrate the tree-level contribution of $\Lambda^{-5}d^c d^c d^c u^c L^i Q^j \epsilon_{ij}$ to $n\to M\nu$, along with our estimate for the partial width.  The loop-level contribution overwhelms the tree-level for $\Lambda^2v/\Lambda_{\rm QCD}^3\gg 16\pi^2$, which is the case for all effective scales $\Lambda$ of interest.

%{\bf For the loop to dominate,  $\Lambda^4v^2/\Lambda_{\rm QCD}^6\gtrsim 10^6$, right?}

According to the procedure described in Sec.~\ref{sec:seven}, our results for the nucleon lifetime and that of $n-\bar{n}$ oscillation are listed in Table~\ref{table_bounds}. The different lifetime estimates depend on the down-type-quark, up-type-quark, or charged-lepton Yukawa couplings $y_d, y_u, y_e$, respectively, and we list only the leading contribution to the different two-body final-state nucleon decays. We further assume that the operators are flavor indifferent and hence assume the up-type-quark (down-type-quark) [charged-lepton] Yukawa coupling to be that of the top-quark (bottom-quark) [tau-lepton]. For the numerical estimates displayed in Table~\ref{table_bounds}, we use $y_u v=120$~GeV and $y_d v=2.5$~GeV in order to acknowledge the renormalization group running between the GUT scale and $\Lambda$. Given the order-of-magnitude nature of our estimates, these choices are perfectly adequate. Our analytic estimate for the $n-\bar{n}$ oscillation rate is also depicted in Table~\ref{table_bounds}, and agrees with similar estimates computed in Ref.~\cite{Rao:1982gt,Mohapatra:2009wp}. Experimentally, the time scale for $n-\bar{n}$ oscillations is constrained to be greater than 2.7~years and 4.1~years at the 90\% confidence level, for free and bound neutrons, respectively~\cite{Beringer:1900zz}.  An improvement of up to three orders of magnitude is expected from proposed next-generation experiments \cite{Babu:2013yww}. However, since $n-\bar{n}$ oscillations are, at leading order, a dimension-nine phenomenon, these results provide much weaker constraints than those provided by nucleon decay searches, modulo large GUT-breaking effects. It is clear that $\mathcal{O}^{\rm GUT}_{5}$-related baryon-number violating processes are many orders of magnitude faster than the current experimental bounds. In the absence of large GUT-breaking effects, we can state that  the physics that yields $\mathcal{O}^{\rm GUT}_{5}$ is disfavored by the non-observation of nucleon decay as the dominant source of the observed neutrino masses. 

All other dimension-nine operators mediate, at low energies, $|\Delta L|=2$, as listed in Table~\ref{app_table}. Of the $|\Delta L|=2$ operators listed in Refs.~\cite{Babu:2001ex,de Gouvea:2007xp}, ${\cal O}_{6,9,10,11_b,12_a,12_b,13,14_a,14_b,15,16,17,18,19,20}$, along with the `singlets' operator $e^c e^c u^c u^c d^{c\dagger}d^{c\dagger}$\footnote{The `singlets' operator accommodates the observed neutrino masses for very small $\Lambda\sim 1$~GeV and, for this reason, it is not considered in Refs.~\cite{Babu:2001ex,de Gouvea:2007xp,Angel:2012ug}.} are consistent with grand unification. ${\cal O}_{5,7,11_a}$ are not, according to our procedure, consistent with grand unification. Among the ``allowed'' $|\Delta L|=2$ operators of mass-dimension nine, only ${\cal O}_{6,11_b,12_a,12_b,14_a,14_b,15,17}$ can be made to fit neutrino data. The remaining operators are entwined with these in such a way that, modulo large GUT-breaking effects, they can only contribute at a subdominant level to the observed neutrino masses. 

The different accessible two-body-final-state nucleon decay modes are listed in  Table~\ref{table_bounds} for all the dimension-nine operators. In the case of $\mathcal{O}^{\rm GUT}_{8_a,8_b}$, proton decay occurs only at the two-loop level -- an extra $W$-boson loop is required in order to ``convert'' a down-quark into an up-quark -- while in the case of $\mathcal{O}^{\rm GUT}_{9}$ the two-body nucleon decay occurs by closing a Higgs boson loop, in such a way that the decay rate is not proportional to any of the Yukawa couplings. In spite of the one-loop suppression, the low (relative to the dimension-seven operators discussed in the previous section) effective scales required to fit the observed neutrino masses translate into shorter nucleon lifetimes. As far as baryon number violating processes are concerned, only ${\cal O}^{\rm GUT}_{7_a}$ and ${\cal O}^{\rm GUT}_{9}$ are not severely constrained, even if one takes into account that ${\cal O}^{\rm GUT}_{8_a}$ requires more than one fermion generation in order to exist. Some operators are ``ruled out'' by many orders of magnitude. This, in turn, means that if grand unification is real the only dimension-nine $|\Delta L=2|$ effective operators allowed by all data, assuming GUT-breaking effects are under control,  are ${\cal O}_{6,12_a}$. These, furthermore, loosely predict that $|\Delta(B-L)|=2$ nucleon decay is to be observed by next-generation searches. 

We expect that new physics that manifests itself at the tree-level via dimension-eleven operators will mediate $|\Delta(B-L)|=2$\footnote{Dimension-eleven operators are still ``too small'' to contain $|\Delta(B-L)|=6$ operators.} nucleon decay at the two-loop level (or higher). We also anticipate, at the one-loop level, $|\Delta B|=2$ processes. In spite of the higher-order nature of the effects, similar to what we observe when comparing dimension-seven and dimension-nine operators, we expect the nucleon lifetimes consistent with the observed neutrino data to be, on average, shorter, given the smaller required $\Lambda$ values. We anticipate, therefore, only a small subset of the dimension-eleven ${\cal O}_{I}$, if any, to be consistent with grand unification, modulo large GUT-breaking effects. For this reason, we did not conduct a detailed study of dimension-eleven ${\cal O}^{\rm GUT}_{J}$ operators.

\section{Summary and Concluding Remarks}

We argue that, if nature is described by a grand-unified gauge theory at very short distances, searches for low-energy processes that violated $B-L$ by two units  are expected to play a nontrivial role in piecing together the neutrino mass puzzle, assuming the neutrinos are Majorana fermions. We estimate, for a large class of LNV scenarios, expected rates for $|\Delta(B-L)|=2$ processes, more concretely two-body-final-state decays of nucleons and the neutron--antineutron oscillation rate. Our results are summarized in Table~\ref{table_bounds}.

Our approach is as follows. We assume the existence of new heavy (mass larger than the weak scale) states $X$ that break lepton-number conservation explicitly at the heavy mass scale. Those, when integrated out, will lead to effective operators suppressed by different powers of $\Lambda$ that mediate $|\Delta L|=2$ phenomena, including non-zero Majorana neutrino masses. The contribution to the neutrino mass from a generic scenario can be estimated by assuming that, at the tree-level, its low-energy effects are captured by a leading $D$-dimensional operator, $D\ge7$, and that the coefficient of the Weinberg operator, which is generated at some loop-level, can be estimated from the leading operator \cite{Babu:2001ex,de Gouvea:2007xp}. Following this procedure, one determines the values of $\Lambda$ that lead to the observed neutrino masses. If grand unification is real, the $X$ fields must be part of larger GUT multiplets $X_5$. The act of integrating out the other components of $X_5$, assuming these are all heavy, will lead, at the tree-level, to other effective operators of the same dimension $D$, all of which violate $B-L$ by two (or six, ten, etc) units, as we demonstrate in Appendix~\ref{operatordim}, and mediate BNV and LNV low-energy phenomena at different loop-levels. As with the $|\Delta L|=2$ leading operator, we can estimate the contributions of these $|\Delta (B-L)|=2$ operators to nucleon decay as a function of $\Lambda$. By assuming that the effective scale $\Lambda$ of GUT-related operators are the same we associate the neutrino mass data to expected rates for $|\Delta(B-L)|=2$ processes. 

In order to define which $D$-dimensional $|\Delta (B-L)|=2$ operators are related, we constructed all $D=7$ and $D=9$ ${\cal O}^{\rm GUT}$ operators made up of the GUT matter fields -- $\psi$, in the $\mathbf{\bar{5}}$ representation and $\chi$, in the $\mathbf{10}$ representation --  and Higgs fields, $\Phi$, in the $\mathbf{\bar5}$ representation. These are listed in Table~\ref{app_table}. The related $D$-dimensional $|\Delta (B-L)|=2$ operators are then simply the different ``components'' of ${\cal O}^{\rm GUT}$ after GUT-breaking. In Table~\ref{app_table} we list which $|\Delta L|=2$ operators are included in the different ${\cal O}^{\rm GUT}$. 

Strictly speaking, this procedure is correct in the case $\Lambda\gg M_{\rm GUT}$, the GUT-breaking scale. If this were the case, at the scale $\Lambda$, the theory would be described by the GUT version of the Standard Model plus $X_5$, while at energy scales around $M_{\rm GUT}$ physical phenomena would be properly described by the GUT version of the Standard Model plus the effective operators ${\cal O}^{\rm GUT}$. Finally, in the infrared, physical phenomena would be properly described by the Standard Model, plus the ``components'' of ${\cal O}^{\rm GUT}$ and other effective operators suppressed by different powers of $M_{\rm GUT}$.\footnote{As we argued earlier, these do not mediate $|\Delta (B-L)|=2,6,10,\ldots$ effects except through their mixing with the $|\Delta (B-L)|$ physics, captured by ${\cal O}^{\rm GUT}$.} The coefficients of the different higher-dimensional operators would be the same, modulo calculable quantum corrections (renormalization-group running between $M_{\rm GUT}$ and the infrared). Instead, we are interested in the case $\Lambda\ll M_{\rm GUT}$ (but still $\Lambda\gg v$, the weak scale). For the purposes of establishing which $D$-dimensional operators generated by the same GUT-connected physics are related, the procedure described above is valid. However, already at the scale $\Lambda$, GUT-breaking effects render the coefficients of the different components of ${\cal O}^{\rm GUT}$ different from one another. These effects are model dependent and hence cannot be calculated following our approach.  

In order to relate the experimentally-constrained neutrino masses to the rates for low-energy baryon-number violating processes, we posit that GUT-breaking effects, along with renormalization-group running effects between $\Lambda$ and the infrared, are small enough that all related  $|\Delta (B-L)|=2$ operators have the same coefficient. While this appears to be a natural minimalistic assumption, it is by no means guaranteed to be correct. Numerically, our estimates are not sensitive to order-one effects, or even effects that lead to one order of magnitude differences. Order-one GUT-breaking effects are known and required in order to render the different Yukawa couplings consistent with low-energy observations. On the other hand, the doublet-triplet splitting problem -- the fact that the color-triplet Higgs field is many orders of magnitude heavier than the Standard Model Higgs doublet, in spite of the fact that the two fields belong to the same GUT multiplet -- indicates that there are circumstances where GUT-breaking effects can be most dramatic (more than eight orders of magnitude).   

At face value, our results indicate that if grand unification is real most lepton-number violating scenarios captured by the $|\Delta L|=2$ ${\cal O}_I$ operators of dimension $D=7$ or 9 that lead to neutrino masses \cite{Babu:2001ex,de Gouvea:2007xp,Angel:2012ug} are inconsistent with either the GUT hypothesis or the non-observation of $|\Delta(B-L)|=2$ nucleon decay. Roughly speaking, the scenarios consistent with current data and the GUT hypothesis are associated to lepton-number breaking scales $\Lambda\gtrsim 10^{10}$~GeV. This trend seems to hint that very few, if any, $D=11$ scenarios are allowed (see Refs.~\cite{Babu:2001ex,de Gouvea:2007xp,Angel:2012ug}). 

The very high $\Lambda$ values raise concerns associated with the stability of the weak scale. The so-called hierarchy problem requires the new lepton-number breaking physics to be very weakly coupled \cite{Casas:2004gh,Farina:2013mla,deGouvea:2014xba} or the presence of new degrees of freedom with masses between $\Lambda$ and the weak scale. This concern, of course, is already present if GUTs are realized in nature, independent from whether or how lepton-number is violated. If nature were supersymmetric at short distances and supersymmetry was broken slightly above the weak scale, the GUT-breaking, LNV, and electroweak breaking scales could all co-exist ``naturally.'' The presence of low-energy supersymmetry, however, could lead to quantitatively different results as far as this work is concerned. Superpartners to the Standard Model fields, for example, allow for more and lower-dimensional $|\Delta(B-L)|=2$ operators. The act of integrating out weak-scale superpartners, in turn, may end up providing more stringent constraints than the ones discussed here.   

We reiterate that our most robust result is to list which $|\Delta(B-L)|=2$ operators are ``related,'' a result which highlights some $|\Delta(B-L)|=2$ observables as especially interesting for understanding nonzero neutrino masses. GUT-breaking effects can, for example, render $|\Delta(B-L)|=2$ nucleon decay processes significantly slower or $|\Delta B|=2$ processes like neutron--antineutron significantly faster. We also assumed that all operators are flavor-indifferent, i.e., all lepton and quark flavors interact with the new degrees of freedom with similar strength. It is important to keep in mind that, in order to fit neutrino masses, no new interactions involving first-generation quarks are required. If this turns out to be the case, none of the bounds discussed here apply.  Finally, we remind the readers that we restricted our discussion to the gauge group $SU(5)$. Larger gauge groups contain more degrees of freedom which may be light and include $B-L$ as a gauged subgroup, both of which render the discussion potentially more complex than and qualitatively different from the one presented here.

\section*{Acknowledgments}

We thank K.S.~Babu for comments on the manuscript. The work of AdG and AK is sponsored in part by the DOE grant \# DE-FG02-91ER40684.  AK is supported in part by the Department of Energy Office of Science Graduate Fellowship Program (DOE SCGF), made possible in part by  the American Recovery and Reinvestment Act of 2009, administered by ORISE-ORAU under contract no.~DE-AC05-06OR23100. J.H.-G. is supported by the MINECO under the FPU program.

\appendix
\section{The Relationship between $\Delta B$, $\Delta L$ and Operator Dimension}
\label{operatordim}
 
A relationship exists between the dimension of an operator in the Standard Model and its $\Delta B$ and $\Delta L$ assignments, as first explored in Ref.~\cite{Rao:1983sd}. To investigate the nature of this relationship, we consider operators constructed only of Standard Model fermions and Higgs fields, i.e., operators with no derivatives nor gauge bosons.  We can count the number of left-handed ($N_Q, N_L, N_u, N_d, N_e$) and right-handed fermions ($N_{Q^{\dagger}}, N_{L^{\dagger}}, N_{u^{\dagger}}, N_{d^{\dagger}}, N_{e^{\dagger}}$) and the number of Higgs fields ($N_H$ and $N_{H^*}$) in each operator. For example, the operator $L\tilde{H}e^c$ has $N_L = N_{H^*}= N_e=1$ and $N_Q = N_u = N_d = N_H =   N_{Q^\dagger} = N_{u^\dagger}  = N_{d^\dagger} = N_{L^\dagger} = N_{e^\dagger} = 0$. $\Delta B$ and $\Delta L$, for a given operator, can be expressed as a function of the different $N$s as
\begin{eqnarray}
\text{Baryon number: } &&   \Delta B = \frac{1}{3}\left(N_Q - N_{Q^\dagger} - N_u + N_{u^\dagger} - N_d + N_{d^\dagger}\right), \label{bnum} \\
\text{Lepton number: } &&   \Delta L = N_L - N_{L^\dagger} - N_e + N_{e^\dagger}. \label{lepnum}
\end{eqnarray}
Hypercharge and Lorentz invariance impose the following constraints:
\begin{eqnarray}
\text{hypercharge invariance: } && 0 = \frac{1}{3}(N_Q - N_{Q^\dagger}) -\frac{4}{3}(N_u-N_{u^\dagger})+ \frac{2}{3}(N_d-N_{d^\dagger})  \label{hyper} \\
&& \hspace{0.35in} - ~  (N_L-N_{L^\dagger}) + 2(N_e-N_{e^\dagger})  + (N_H - N_{H^*}), \nonumber \\
\text{Lorentz invariance: }  &&0,2,4,... = N_d + N_u + N_Q + N_L + N_e, \\
\text{Lorentz invariance: }  &&0,2,4,... = N_{d^\dagger} + N_{u^\dagger} + N_{Q^\dagger} + N_{L^\dagger} + N_{e^\dagger}, \label{lorentzRH}
\end{eqnarray}
while the dimension of the operator, $D$, is defined as
\begin{equation}
D = \frac{3}{2}\left(N_Q+N_{Q^\dagger} + N_u + N_{u^\dagger} + N_d + N_{d^\dagger} + N_L + N_{L^\dagger} + N_e + N_{e^\dagger}  \right) + N_H + N_{H^*}.
\end{equation}
The insertion of Eqs.~(\ref{bnum}) and (\ref{lepnum}) into the definition of $D$ yields
\begin{equation}
\label{D1}
D = 3\left(N_d + N_e + N_u + N_{Q^\dagger} + N_{L^\dagger} + \frac{3}{2}   \Delta B + \frac{1}{2}   \Delta L\right) + N_H + N_{H^*}.
\end{equation}
Eqs.~(\ref{bnum}) and (\ref{lepnum})  can also be inserted into Eq.~(\ref{hyper}) to yield
\begin{equation}
\label{ndne}
N_d + N_e = N_d^\dagger + N_u - N_u^\dagger + N_e^\dagger - N_H + N_H^\dagger -    \Delta B +  \Delta L.
\end{equation}
From here, Eq.~(\ref{ndne}) can be substituted into Eq.~(\ref{D1}):
\begin{equation}
D = 3\left(2N_u + N_{d^\dagger} - N_{u^\dagger} + N_{e^\dagger} + N_{Q^\dagger} + N_{L^\dagger} + \frac{1}{2}  \Delta B + \frac{3}{2}   \Delta L \right) - 2N_H + 4N_{H^*}.
\end{equation}
From Eq.~(\ref{lorentzRH}), one can note that $N_{d^\dagger} - N_{u^\dagger} + N_{Q^\dagger} + N_{L^\dagger} + N_{e^\dagger} = (0,2,4...) - 2N_{u^\dagger}$ is an even number.  Therefore
\begin{equation}
\left| \frac{1}{2}   \Delta B + \frac{3}{2}  \Delta L\right| \in \mathbb{N} \left\{ \begin{array}{ll} \text{odd} &\leftrightarrow \text{$D$ is odd,} \\ \text{even} &\leftrightarrow \text{$D$ is even.} \end{array} \right.
\end{equation}
This relationship between $\Delta B$, $\Delta L$, and the dimension of the operator, $D$, only assumes hypercharge and Lorentz invariance. It is straightforward to show that the relation remains unchanged if right-handed Standard Model gauge singlet neutrinos with lepton number $+1$ (or left-handed gauge-singlet antineutrinos with lepton number $-1$) are introduced.

 \end{document}